\documentclass[aps,showpacs,prep,graphics]{revtex4}
\usepackage{amssymb}
\usepackage[dvips]{graphicx}
\usepackage{amsmath}
\usepackage{bm}
\usepackage{epsfig}
\usepackage{graphicx}
\usepackage{caption}
\usepackage{wrapfig}
\usepackage{color}

\begin{document}

\title{Gravitoelectromagnetic inflation and seeds of cosmic magnetic fields  from geometrical Weyl-invariant scalar-tensor theory of gravity }

\author{  $^{1}$ M. Montes, $^{2}$ Jos\'e Edgar Madriz Aguilar\thanks{E-mail address: madriz@mdp.edu.ar} and $^{3}$ V. Granados
\thanks{E-mail address: mariana.montnav@gmail.com} }
\affiliation{$^{1}$ Departamento de Ciencias Naturales y Exactas \\ Centro Universtario de los Valles\\
Carretera Guadalajara-Ameca Km. 45.5, C.P. 46600, Ameca, Jalisco, M\'exico.\\
and\\ 
$^{2}$ Departamento de Matem\'aticas, Centro Universitario de Ciencias Exactas e ingenier\'{i}as (CUCEI),
Universidad de Guadalajara (UdG), Av. Revoluci\'on 1500 S.R. 44430, Guadalajara, Jalisco, M\'exico.  \\
and\\
$^{3}$ Departamento de F\'isica, Centro Universitario de Ciencias Exactas e ingenier\'{i}as (CUCEI),
Universidad de Guadalajara (UdG), Av. Revoluci\'on 1500 S.R. 44430, Guadalajara, Jalisco, M\'exico.  \\
E-mail:  jose.madriz@academicos.udg.mx, 
madriz@mdp.edu.ar,
mariana.montes@academicos.udg.mx, cromero@fisica.ufpb.br, victor.granados@alumnos.udg.mx}

\begin{abstract}
We investigate cosmological inflationary scenarios from a  gravitoelectromagnetic theory. Our work is formulated in the light of a recently introduced geometrical Weyl-Invariant scalar-tensor theory of gravity, where the nature of both the electromagnetic potential and the inflaton field is attributed to the space-time geometry. We obtain a Harrison-Zeldovich power spectrum for quantum fluctuations of the inflaton field. In our model the electromagnetic fields have also a nearly scale invariant power spectrum for a power-law inflation. We found that the the seed magnetic fields have a nearly scale invariant power spectrum and generate in the present times cosmic magnetic fields of the order  $\lesssim 10^{9}$ gauss, in good  agreement with CMB observations.
\end{abstract}

\maketitle

\vskip .5cm
 Weyl-Integrable geometry, gravitoelectromagnetism, scalar-tensor gravity, inflation, cosmic magnetic fields.

\section{Introduction}

Inflation is a period of accelerated expansion in the evolution of the early universe that has emerged as an attempt to solve the problems of the Big Bang cosmology. Observational data of Cosmic Microwave Background (CMB) anisotropies give evidence of this epoch. The quantum fluctuations of the inflaton field  originate the seeds of large scale structure of the universe \cite{In1,In2,In3,In4}.  In the large scale structure formation an explanation for the cosmic magnetic fields is one of the challenging problems.  One of the most accepted ideas consists in that a seed of a cosmic magnetic  field sufers an exponentiation due to a galactic dynamo mechanism \cite{In5, In6}. It is assumed that this mechanism could happend during inflation \cite{In5}. However, to have a magnetic seed during inflation it is necessary  a gravity model with an electromagnetic contribution. Models of gravitoelectromagnetism are theoretical settings capable to address this problem in a natural manner. In the literature we can find a variety of them with different formulations (the reader might see for example \cite{In7,In8,In9,In10,In11}).\\

Recently, a new geometrical approach of the class of scalar-tensor theories of gravity has been introduced \cite{In12,In13}. The main idea is that if a Palatini variational principle is adopted the natural background geometry of this class of theories is determined by the Weyl-integrable compatibility condition \cite{In12}. The requirement that the action and the compatibility condition must have the same symmetry group led to the need to introduce a new action that would satisfy such a requirement of invariance. However, the Weyl invariance is achieved when a gauge geometrical vector field is introduced in the covariant derivative in a similar manner as it is usually done in quantum gauge theories\cite{In14}. In the so called Einstein-Riemann frame it is obtained a theory of gravitoelectromagnetism where both the scalar field and the electromagnetic potential have a geometrical origin \cite{In14}. Several topics have been address in the framework of this geometrical scalar-tensor gravity among them we can count cosmological models \cite{WCR3,WCRA1}, scalar fluctuations of the metric during inflation \cite{WCRA2}, the singularity issue \cite{WCRA3} and some others \cite{WCRA4}.\\

In this letter we develop a cosmological inflationary formalism in which seeds of electromagnetic fields can be generated. The paper is organized as follows. In section I we give a little introduction. In section II we obtain the field equations corresponding to the model of gravitoelectromagnetism. In section III we formulate the dynamical equations that govern both the inflaton and electromagnetic fluctuations during inflation. In section IV, as an application of the previous formalism we study a power-law inlfationary model. Finally, section V is for some final comments.

\section {The geometrical Weyl-Invariant Gravity Theory}

 We start by considering a scalar-tensor theory of gravity in vacuum described by the action
\begin{equation}\label{f1}
S=\frac{1}{16\pi}\int d^{4}x\sqrt{-g}\left\lbrace \Phi {\cal R}+\frac{\tilde{\omega}(\Phi)}{\Phi}g^{\mu\nu}\Phi_{,\mu}\Phi_{,\nu}-\tilde{V}(\Phi)\right\rbrace ,
\end{equation}
where ${\cal R}$ denotes the Ricci scalar, $\tilde{\omega}(\Phi)$ is a function of the scalar field $\Phi$ and $\tilde{V}(\Phi)$ is a scalar potential. Introducing $\varphi=-\ln (G\Phi)$ the equation (\ref{f1}) reads \cite{WCRA2}
\begin{equation}\label{f2}
S=\int d^{4}x\sqrt{-g}\left\lbrace e^{-\varphi}\left[\frac{{\cal R}}{16\pi G}+\frac{1}{2}\omega (\varphi)g^{\mu\nu}\varphi_{,\mu}\varphi_{,\nu}\right]-V(\varphi)\right\rbrace,
\end{equation}
where we have made the identifications $(1/2)\omega(\varphi)=(16\pi G)^{-1}\tilde{\omega}[\varphi(\Phi)]$ and $V(\varphi)=(16\pi)^{-1}\tilde{V}(\varphi(\Phi))$. Adopting a  Palatini variational procedure we obtain the non-metricity condition \cite{In12}
\begin{equation}\label{f3}
\nabla _{\mu}g_{\alpha\beta}=\varphi_{,\mu}g_{\alpha\beta}.
\end{equation}
As the Palatini variational principle is used to determine the background geometry for a given action \cite{WCR1,WCR2,WCR3}, the equation (\ref{f3}) can be interpreted as the Weyl-Integrable geometry is the background geometry for the action (\ref{f2}). It is a well-known property that (\ref{f3}) is invariant under the Weyl group of transformations \cite{In12}
\begin{eqnarray}\label{f4}
\bar{g}_{\alpha\beta}&=&e^{f}g_{\alpha\beta}\\
\label{f5}
\bar{\varphi}&=&\varphi +f,
\end{eqnarray}
where $f=f(x^{\alpha})$ is a well-behaved function of the spacetime coordinates. The invariance of (\ref{f3}) is achieve when the both transformations (\ref{f4}) and (\ref{f5}) are applied at the same time. In fact a pure conformal transformation does not preserve (\ref{f4}).   As an action is by definition a scalar, in this case under the Weyl group, it is natural to require the invariance of the action (\ref{f2}) under the Weyl transformations (\ref{f4})-(\ref{f5}). However, it is not difficult to see that the  kinetic term does not satisfy such requirement. Thus, an invariant action results to be \cite{In14}
\begin{equation}
{\cal S}=\int d^{4}x\sqrt{-g}\,e^{-\varphi}\left[\frac{{\cal R}}{16\pi G}+\frac{1}{2}\omega(\varphi)g^{\alpha\beta}\varphi_{:\alpha}\varphi_{:\beta}-V(\varphi)e^{-\varphi}-\frac{1}{4}H_{\alpha\beta}H^{\alpha\beta}e^{-\varphi}\right],\label{f14}
\end{equation}
where $\varphi_{:\mu}=(\,^{(w)}\nabla_{\mu}+\gamma B_{\mu})\varphi,$ is a gauge covariant derivative, $H_{\alpha\beta}=W_{\beta ,\alpha}-W_{\alpha ,\beta}$ with $W_{\alpha}=\varphi B_{\alpha}$, and the transformation rules 
\begin{eqnarray}\label{f10a}
\bar{\varphi}\bar{B}_{\mu} &=& \varphi B_{\mu}-\gamma^{-1}f_{,\mu},\\
\bar{\omega}(\bar{\varphi})&=&\omega(\bar{\varphi}-f)=\omega(\varphi),\label{f10b}\\
\bar{V}(\varphi)&=& V(\bar{\varphi}-f)=V(\varphi),\label{f10c}
\end{eqnarray}
must be valid. From now on $^{(w)}\nabla_{\mu}$ will denote the covariant derivative calculated with the Weyl connection. The set of transformations (\ref{f4}), (\ref{f5}) and (\ref{f10a}) can be interpreted as they lead from the Weyl frame $(M,g,\varphi,B_{\mu})$ to a new Weyl frame $(M,\bar{g},\bar{\varphi},\bar{B}_{\mu})$ with the same causal structure and preserving geodesics \cite{WCR3}. By choosing $f=-\varphi$ in (\ref{f3}) the Riemannian compatibility condition $\nabla_{\alpha}h_{\mu\nu}=0$ is recovered for the effective metric  $h_{\mu\nu}=\bar{g}_{\mu\nu}=e^{-\varphi}g_{\mu\nu}$. The frame $(M,h,0,A_{\mu}=\bar{B}_{\mu})$ is known as the Riemann frame. 

\section{Gravitoelectromagnetism from geometrical Weyl-Invariant gravity}

We shall now derive a gravitoelectromagnetic theory from the previous geometrical formalism. In the Riemann frame the Weyl-invariant action (\ref{f14}) reads
\begin{equation}\label{Rie2}
{\cal S}=\int d^{4}x\sqrt{-h}\left[\frac{R}{16\pi G}+\frac{1}{2}\omega(\phi)h^{\alpha\beta}{\cal D}_{\alpha}\phi{\cal D}_{\beta}\phi-V(\phi)-\frac{1}{4}F_{\alpha\beta}F^{\alpha\beta}\right],
\end{equation}
where ${\cal D}_{\mu}=\nabla_{\mu}+\gamma A_{\mu}$,  the operator $\nabla_{\lambda}$ is denoting the Riemannian covariant derivative, $\phi$ is the residual scalar field $\varphi$ as view in the Riemann frame and $F_{\mu\nu}=A_{\nu ,\mu}-A_{\mu,\nu}$.  For the action (\ref{Rie2}) can be formally identical to an Einstein-Hilbert action for a scalar field minimally coupled to gravity with non-canonical kinetic term and a gauge electromagnetic field $A_\mu$,  the next gauge transformation must be valid  
\begin{equation}\label{yq1}
\overset{\smile}{A}_{\mu}=A_{\mu}-\gamma^{-1}\sigma_{,\mu},
\end{equation}
where $\sigma =\sigma (x^{\alpha})$. It is important recall that unlike other theories of gravitoelectromagnetism, in our formalism $A_{\mu}$ has been introduced as a consequence of imposing a geometrical symmetry (the Weyl-Invariance of (\ref{f14})), and not only by requiring a gauge symmetry. To ensure the invariance of the action (\ref{Rie2}) under  (\ref{yq1}), the next internal symmetries must also hold
\begin{eqnarray}\label{yq2}
\overset{\smile}{\phi}&=&\phi e^{\sigma},\\
\label{yq2p}
\overset{\smile}{\omega}(\overset{\smile}{\phi})&\equiv& e^{-2\sigma}\omega(e^{-\sigma}\overset{\smile}{\phi})=\omega(\phi)\\
\label{yq2q}
\overset{\smile}{V}(\overset{\smile}{\phi})&\equiv& V(e^{-\sigma}\overset{\smile}{\phi})=V(\phi).
\end{eqnarray}
The action (\ref{Rie2}) can be extended by adding a source term for $A_{\mu}$ and a usual Feymann gauge term in the form 
\begin{equation}\label{yq3}
{\cal S}=\int d^{4}x\sqrt{-h}\left[\frac{R}{16\pi G}+\frac{1}{2}\omega(\phi)h^{\alpha\beta}{\cal D}_{\alpha}\phi{\cal D}_{\beta}\phi-V(\phi)-\frac{1}{4}F_{\alpha\beta}F^{\alpha\beta}-\frac{\lambda}{2}(\nabla_{\alpha}A^{\alpha})^2-J^{\alpha}A_{\alpha}\right],
\end{equation}
where $J^{\mu}$ is a conserved current density. The field equations obtained from the action (\ref{yq3}) are
\begin{eqnarray}\label{Rie6}
&& G_{\mu\nu}=-8\pi G \left[\omega(\phi){\cal D}_{\mu}\phi{\cal D}_{\nu}\phi-\frac{1}{2}h_{\mu\nu}\left(\omega(\phi)h^{\alpha\beta}{\cal D}_{\alpha}\phi{\cal D}_{\beta}\phi- 2V(\phi)\right)-\tau_{\mu\nu}^{(em)}\right]\\
&& \omega(\phi)\Box\phi+\frac{1}{2}\omega^{\prime}(\phi)h^{\mu\nu}{\cal D}_{\mu}\phi{\cal D}_{\nu}\phi-\gamma\omega^{\prime}(\phi)A^{\mu}\phi{\cal D}_{\mu}\phi+\gamma\omega(\phi)\nabla_{\mu}A^{\mu}-\gamma^2\omega(\phi)A^{\mu}A_{\mu}\phi+V^{\prime}(\phi)=0,\label{Rie7}\\
&& \nabla_{\mu}F^{\mu\nu}+\lambda \nabla^{\nu}(\nabla_{\mu}A^{\mu})= J^{\nu}-\gamma\omega(\phi)h^{\mu\nu}\phi{\cal D}_{\mu}\phi,\label{Rie8}
\end{eqnarray}
with  $\Box =h^{\mu\nu}\nabla_{\mu}\nabla_{\nu}$ being the usual D'Alambertian operator, $\tau_{\mu\nu}^{(em)}=T_{\mu\nu}^{(em)}-h_{\mu\nu}J^{\alpha}\! A_{\alpha}$, with $T_{\mu\nu}^{(em)}=F_{\nu\beta}F_{\mu}\,^{\beta}-\frac{1}{4}h_{\mu\nu}F_{\alpha\beta}F^{\alpha\beta}$ being the energy-momentum tensor for a free electromagnetic field. The field equations (\ref{Rie6}), (\ref{Rie7}) and (\ref{Rie8}) formally describe a theory of gravitoelectromagnetism on an effective Riemannian geometrical background. In the next section we shall consider cosmological applications of these field equations. In particular we will focus in how to derive inflationary models in which $\phi$ can play the role of the inflaton field.

\section{Inflatoelectromagnetic Fluctuations}

In order to develop an inflationary model from this formalism let us consider a spatially flat Friedmann-Robertson-Walker metric in the form
\begin{equation}\label{inff7}
ds^2=dt^2-a^2(t)(dx^2+dy^2+dz^2),
\end{equation}
with $a(t)$ being the usual cosmological scale factor. The cosmological principle allow us to assume that the inflaton scalar field and the electromagnetic potential can be written in the form
\begin{equation}\label{inff1}
\phi(x^{\lambda})=\phi_{c}(t)+\delta\phi(x^{\lambda}),\quad A^{\mu}(x^{\sigma})=\delta A^{\mu}(x^{\sigma}),
\end{equation}
where $\phi_{c}(t)=<\phi(x^{\lambda})>$, $<\delta\phi>=<\delta\dot{\phi}>=0$ and $<A^{\mu}(x^{\sigma})>=<\delta A^{\mu}(x^{\sigma})>=0$. Here $\delta\phi$ denotes the quantum fluctuations of the inflaton scalar field and $\delta A_{\alpha}$ accounts for the quantum fluctuations of the electromagnetic field. Hence it follows from the equations (\ref{Rie7}), (\ref{Rie8}) and (\ref{inff7}) that the classical and quantum parts for the inflaton field, respectively, for $J^{\mu}=0$, can be written as
\begin{eqnarray}
&& \ddot{\phi}_c + 3(1+\gamma)H\dot{\phi}_c+\frac{\omega^{\prime}(\phi_c)}{2\omega(\phi_c)}\dot{\phi}_c^2+\frac{V^{\prime}(\phi_c)}{\omega(\phi_c)}=0,\label{inff2}\\
&& \ddot{\delta\phi}+\left(3H+\frac{\omega^{\prime}(\phi_c)}{\omega(\phi_c)}\dot{\phi}_c\right)\dot{\delta\phi}-\frac{1}{a^2}\nabla^{2}\delta\phi + \frac{1}{\omega(\phi_c)}\left[V^{\prime\prime}(\phi_c)+\omega^{\prime}(\phi_c)(\ddot{\phi}_c+3H\dot{\phi}_c)+\frac{1}{2}\omega^{\prime\prime}(\phi_c)\dot{\phi}_c^2\right]\delta\phi=0.\label{inff3}
\end{eqnarray}
It is not difficult to show that in a similar manner from the equation (\ref{Rie8}), taking $J^{\mu}=0$, the quantum electromagnetic fluctuations $\delta A_{\mu}$ are determined by 
\begin{equation}
 \nabla_{\mu}\delta F^{\mu\nu}+\lambda\nabla^{\nu}(\nabla_{\mu}\delta A^{\mu})=\delta J^{\nu}-\gamma\omega(\phi_c)h^{\mu\nu}\left[\gamma\phi^2_c\delta A_{\mu}+\phi_c\partial_{\mu}\delta\phi+(\partial_{\mu}\phi_c)\delta\phi\right]-\gamma\omega^{\prime}(\phi_c)h^{\mu\nu}\phi_c\partial_{\mu}\phi_c\delta\phi.\label{inff5}
\end{equation}
The classical part of (\ref{Rie6}) according to (\ref{inff7}) leads to the Friedmann equation
\begin{equation}\label{inff6}
H^2=\frac{8\pi G}{3}\left[\frac{1}{2}\omega(\phi_c)\dot{\phi}^2+V(\phi_c)\right].
\end{equation}
Thus, the classical part of the inflaton field is given by the equations (\ref{inff2}) and (\ref{inff6}), whereas their quantum fluctuations are governed by the expression (\ref{inff3}). The equation (\ref{inff5}) determines the dynamics of $\delta A_{\mu}$ and establishes a relation between the electromagnetic fluctuations $\delta A_{\mu}$ and the ones of the inflaton $\delta\phi$. \\

Now, following a standard quantization procedure, the commutation relation for $\delta\phi$ and its canonical conjugate momentum $\Pi^{0}_{(\delta\phi)}=\frac{\partial L}{\partial \dot{\delta\phi}}$ is given by
\begin{equation}\label{inffn1}
\left[\delta\phi(t,\bar{x}),\Pi^{0}_{(\delta\phi)}(t,\bar{x}^{\prime})\right]=i\delta^{(3)}(\bar{x}-\bar{x}^{\prime}).
\end{equation}
Thus, it follows from (\ref{yq3}) that $\Pi^{0}_{(\delta\phi)}=\omega(\phi_c)\sqrt{-h}(\dot{\delta\phi}+\gamma A_0\delta\phi)$ and then the commutator (\ref{inffn1}) reads
\begin{equation}\label{inffn2}
\left[\delta\phi(t,\bar{x}),\dot{\delta\phi}(t,\bar{x}^{\prime})\right]=\frac{i}{\omega(\phi_c)\sqrt{-h}}\,\delta^{(3)}(\bar{x}-\bar{x}^{\prime}).
\end{equation}

Employing the auxiliary field $\delta\chi$ defined by
\begin{equation}\label{inff14}
\delta\phi(t,\bar{x})=\exp\left[-\frac{3}{2}\int \left(H(t)+\frac{1}{3}\frac{\dot{\omega}(\phi_c)}{\omega(\phi_c)}\right)dt\right]\delta\chi(t,\bar{x}),
\end{equation}
and considering the Fourier expansion 
\begin{equation}\label{inff15}
\delta\chi(t,\bar{x})=\frac{1}{(2\pi)^{3/2}}\int d^{3}k\left[a_{k}e^{i\bar{k}\cdot\bar{x}}\xi_{k}(t)+a_{k}^{\dagger}e^{-i\bar{k}\cdot\bar{x}}\xi_{k}^{*}(t)\right],
\end{equation}
with the asterisk mark denoting complex conjugate, $a_{k}$ and $a_{k}^{\dagger}$ being the annihilation and creation operators, the quantum modes $\xi_k(t)$ according to (\ref{inff3}) are determined by the equation
\begin{equation}\label{inff16}
\ddot{\xi}_k+\left[\frac{k^2}{a^2}-\frac{3}{2}\dot{H}+\frac{1}{2}\frac{\omega^{\prime}(\phi_c)}{\omega(\phi_c)}\ddot{\phi}_c+\frac{1}{2}\left(\frac{\omega^{\prime}(\phi_c)}{\omega(\phi_c)}\right)^2\dot{\phi}_c^{2}-\frac{9}{4}\left(H+\frac{\omega^{\prime}(\phi_c)}{3\omega(\phi_c)}\dot{\phi}_c\right)^2+3H\frac{\omega^{\prime}(\phi_c)}{\omega(\phi_c)}\dot{\phi}_c+\frac{V^{\prime\prime}(\phi_c)}{\omega(\phi_c)}\right]\xi_k=0
\end{equation}
The annihilation and creation operators obey the algebra
\begin{equation}\label{inff17}
\left[a_{k},a^{\dagger}_{k^{\prime}}\right]=i\delta^{(3)}(\bar{k}-\bar{k}^{\prime}),\qquad \left[a_{k},a_{k^{\prime}}\right]=\left[a^{\dagger}_{k},a^{\dagger}_{k^{\prime}}\right]=0.
\end{equation}
According to (\ref{inffn2}), (\ref{inff14}) and (\ref{inff15}) the modes must satisfy the normalization condition
\begin{equation}\label{inff18}
\dot{\xi}_{k}^{*}\xi_k-\xi_k^{*}\dot{\xi}_k=\frac{i}{\omega(\phi_c)\sqrt{-h}}\exp \left[3\int \left(H(t)+\frac{1}{3}\frac{\dot{\omega}(\phi_c)}{\omega(\phi_c)}\right)dt\right].
\end{equation}
The squared quantum fluctuations of $\delta\phi$ in the IR-sector (cosmological large scales) are given by
\begin{equation}\label{inff19}
\left<\delta\phi^2\right>_{IR}=\frac{1}{2\pi^2}\exp \left[-3\int \left(H(t)+\frac{1}{3}\frac{\dot{\omega}(\phi_c)}{\omega(\phi_c)}\right)dt\right]\int_{0}^{\epsilon k_H}\frac{dk}{k}k^3\left.\left[\xi_k(t)\xi_k^{*}(t)\right]\right|_{IR},
\end{equation}
where $\epsilon=k_{max}^{IR}/k_p\ll 1$ is a dimensionless parameter with $k_{max}^{IR}=k_{H}(t_r)$ being the wave number related to the Hubble radius at the time $t_r$, which is the time when the modes re-enter to the horizon and $k_p$ is the Planckian wave number. It is well-known that for a Hubble parameter $H=0.5\times 10^{-9}\,M_p$, the values of $\epsilon $ range between $10^{-5}$ and $10^{-8}$, and this corresponds to a number of e-foldings at the end of inflation $N_e=63$.\\

Now, it is not difficult to verify that in the decoupling gauge election
\begin{equation}\label{inff20}
\nabla^{\nu}(\nabla_{\mu}\delta A^{\mu})=\gamma\omega^2(\phi_c)\phi_c\nabla^{\nu}\delta\phi+\left(\gamma\omega^{2}(\phi_c)\partial^{\nu}\phi_c+\gamma\omega^{\prime}(\phi_c)\phi_c\partial^{\nu}\phi_c\right)\delta\phi,
\end{equation}
the equation (\ref{inff5})  becomes
\begin{equation}\label{inff21}
\Box \delta A^{\nu}+\gamma^2\omega(\phi_c)\phi_c^2\delta A^{\nu}=0.
\end{equation}
This equation governs the electromagnetic fluctuations during inflation. In the FRW metric it reads
\begin{equation}\label{inff22}
\ddot{\delta A}^{\nu}+3H\dot{\delta A}^{\nu}-\frac{1}{a^2}\nabla^{2}\delta A^{\nu}+\gamma^2\omega(\phi_c)\phi_c^2\delta A^{\nu}=0.
\end{equation}
In terms of the auxiliary field $\delta\Gamma^{\mu}$ defined by
\begin{equation}\label{inff23}
\delta A^{\mu}(\bar{x},t)=exp\left(-\frac{3}{2}\int H dt\right)\delta\Gamma^{\mu}(\bar{x},t),
\end{equation}
the equation (\ref{inff22}) yields
\begin{equation}\label{inff24}
\ddot{\delta\Gamma}^{\mu}-\frac{1}{a^2}\nabla^{2}\delta\Gamma^{\mu}+\left(-\frac{3}{2}\dot{H}-\frac{9}{4}H^2+\gamma^2\omega(\phi_c)\phi_c^2\right)\delta\Gamma^{\mu}=0.
\end{equation} 
Now, following a covariant quantization procedure the commutation relations for $\delta A^{\mu}$ and $\bar{\Pi}^{\mu}=\frac{\partial L}{\partial(\delta A_{\mu ,t})}$ are given by 
\begin{eqnarray}\label{inff25}
\left[\delta A_{\mu}(\bar{x},t),\bar{\Pi}_{\nu}(\bar{x}^{\prime},t)\right]=ih_{\mu\nu}\delta^{(3)}(\bar{x}-\bar{x}^{\prime}),\\
\label{inff26}
\left[\delta A_{\mu}(\bar{x},t),\delta A_{\nu}(\bar{x}^{\prime},t)\right]=\left[\bar{\Pi}_{\mu}(\bar{x},t),\bar{\Pi}_{\nu}(\bar{x}^{\prime},t)\right]=0.
\end{eqnarray}
Thus, $\bar{\Pi}^{\mu}=\sqrt{-h}\,[\delta F^{\mu 0}-h^{\mu 0}(\nabla_{\sigma}\delta A^{\sigma})]$ according to (\ref{yq3}), therefore  the commutators (\ref{inff25}) and (\ref{inff26}) imply that 
\begin{equation}\label{inff27}
\left[\dot{\delta A}^{\mu}(\bar{x},t),\delta A^{\nu}(\bar{x}^{\prime},t)\right]=\frac{i}{\sqrt{-h}}h^{\mu\nu}\delta^{(3)}(\bar{x}-\bar{x}^{\prime}).
\end{equation}
The auxiliary field $\delta\Gamma^{\alpha}$ can be expressed as a Fourier expansion of the form
\begin{equation}\label{inff28}
\delta\Gamma^{\mu}(\bar{x},t)=\frac{1}{(2\pi)^{3/2}}\int d^{3}\kappa \sum_{\alpha=0}^{3}\epsilon_{(\alpha)}^{\mu}\left[b_{\kappa}^{(\alpha)}e^{i\bar{\kappa}\cdot\bar{x}}\zeta_{\kappa}(t)+b_{\kappa}^{(\alpha)\,\dagger}e^{-i\bar{\kappa}\cdot\bar{x}}\zeta_{\kappa}^{*}(t)\right],
\end{equation}
where the creation and annihilation operators $b_{k}^{(\alpha)\,\dagger}$ and $b_{k}^{(\alpha)}$, respectively, obey the algebra
\begin{equation}\label{inff29}
\left[b_{\kappa}^{(\alpha)},b_{\kappa^{\prime}}^{(\alpha^{\prime})\,\dagger}\right]=-h^{\alpha\alpha^{\prime}}\delta^{(3)}(\bar{\kappa}-\bar{\kappa}^{\prime}),\qquad 
\left[b_{\kappa}^{(\alpha)},b_{\kappa^{\prime}}^{(\alpha^{\prime})}\right]=\left[b_{\kappa}^{(\alpha)\,\dagger},b_{\kappa^{\prime}}^{(\alpha^{\prime})\,\dagger}\right]=0,
\end{equation}
and the four polarisation $4-$vectors $\epsilon_{(\alpha)}^{\mu}$ satisfy $\epsilon_{(\alpha)}\cdot\epsilon_{(\alpha^{\prime})}=h_{\alpha\alpha^{\prime}}$. With the help of (\ref{inff24}) and (\ref{inff28}) the dynamics of the modes is determined by
\begin{equation}\label{inff30}
\ddot{\zeta}_{\kappa}+\left(\frac{\kappa^2}{a^2}-\frac{3}{2}\dot{H}-\frac{9}{4}H^2+\gamma^2\omega(\phi_c)\phi_c^2\right)\zeta_{\kappa}=0.
\end{equation}
The formula (\ref{inff23}) inserted in the equation (\ref{inff27}) generates 
\begin{equation}\label{inff31}
\left[\dot{\delta\Gamma}^{\mu}(\bar{x,t}),\delta\Gamma^{\nu}(\bar{x}^{\prime},t)\right]=\frac{i}{\sqrt{-h}}e^{3\int H dt}\,h^{\mu\nu}\delta^{(3)}(\bar{x}-\bar{x}^{\prime}).
\end{equation} 
Hence the quantum modes $\zeta_{\kappa}(t)$ must obey
\begin{equation}\label{inff32}
\dot{\zeta}_{\kappa}^{*}\zeta_{\kappa}-\zeta_{\kappa}^{*}\dot{\zeta}_{\kappa}=\frac{i}{\sqrt{-h}}\,e^{3\int H dt}.
\end{equation}
The squared fluctuations on cosmological scales for the electromagnetic field $<0|\delta A^{\lambda}\delta A_{\lambda}|0>=<\delta A^2>$ are given by
\begin{equation}\label{inff33}
\left< \delta A^2\right>_{IR}=\frac{e^{-3\int H dt}}{2\pi^2}\int_{0}^{\vartheta \kappa_{H}} \frac{d\kappa}{\kappa}\,\kappa^3 (\zeta_{\kappa}\zeta_{\kappa}^{*})|_{IR},
\end{equation}
where $\vartheta=\frac{\kappa_{max}^{(IR)}}{\kappa}\ll 1$ is a dimensionless parameter. In this formulae $\kappa_{max}^{(IR)}=\kappa_{H}(t_r)$ is the wavenumber related to the Hubble radius at the time $t_r$ which is the time when the modes re-enter to the horizon. $\kappa_p$ is denoting the Planckian wavenumber. Now we are in position of considering as an example a particular inflationary model.

\section{An Example: Power-law inflation}

In this section we shall consider as an illustration of the formalism previously developed a model of power-law inflation. Let us start with the classical dynamics of the inflaton field. Thus, it follows from the equations (\ref{inff2}) and (\ref{inff6}) that
\begin{equation}\label{inff34}
\dot{\phi}_c=-\frac{1}{4\pi G(1+\gamma)}\frac{H^{\prime}(\phi_c)}{\omega(\phi_c)}.
\end{equation}
Hence, inserting (\ref{inff34}) in (\ref{inff6}) we obtain
\begin{equation}\label{inff35}
V(\phi_c)=\frac{3H^2}{8\pi G}-\frac{1}{2(4\pi G)^2(1+\gamma)^2}\frac{H^{\prime}\,^2}{\omega(\phi_c)}.
\end{equation}
Solving (\ref{inff34}) for $H(t)=p/t$ and $\omega(t)=\omega_e(t/t_e)^q$ with $\omega_e=\omega(t_e)$ being $t_e$ the time when inflation ends, we arrive to
\begin{equation}\label{inff36}
\phi_c(t)=\phi_e-\frac{1}{b}(t^{-q/2}-t_e^{-q/2}),
\end{equation}
where $\phi_e=\phi(t_e)$ and $b^2=\frac{4}{q^2}\left(\frac{M_p^2}{4\pi(1+\gamma)}\frac{pt_e^q}{\omega_e}\right)$. Thus, with the help of (\ref{inff35}) and (\ref{inff36}) the scalar potential for the classical part of the inflaton field results to be
\begin{equation}\label{inff37}
V(\phi_c)=\alpha_e p^2(t_e^{-q/2}-b\phi_c)^{4/q},
\end{equation}
where $\alpha_e=\frac{3M_p^2}{8\pi}-\frac{M_p^4t_e^qb^2}{8\pi^2(1+\gamma)^2\omega_eq^2}$. The equation for the modes (\ref{inff16}) becomes
\begin{equation}\label{inff38}
\ddot{\xi}_k(t)+\left(\frac{k^2t_e^{2p}}{a_e^2t^{2p}}+\frac{\beta}{t^2}\right)\xi_k(t)=0,
\end{equation}
where
\begin{equation}\label{inff39}
\beta=\frac{3p}{2}-\frac{3q^2}{4}+\frac{q}{2}-\frac{qp^2}{4}-\frac{q^2p}{6}-\frac{q^3}{36}+\frac{16\alpha_e p^2b^2t_e^q}{\omega_e q^2}-\frac{4\alpha_ep^2b^2t_e^q}{\omega_eq}+3pq.
\end{equation}
The general solution of (\ref{inff38}) can be written in the form
\begin{equation}\label{inff40}
\xi_k(t)=A_1\sqrt{t}\,{\cal H}_{\nu}^{(1)}[x(t)]+A_2\sqrt{t}\,{\cal H}_{\nu}^{(2)}[x(t)],
\end{equation}
where $\nu=\frac{\sqrt{1-4\beta}}{2(1-p)}$ and $x(t)=\frac{t_e^p}{a_e(p-1)}\,kt^{1-p}$. In this equation ${\cal H}_{\nu}^{(1)}$ and ${\cal H}_{\nu}^{(2)}$ are the first and second kind Hankel functions. Considering a Bunch-Davies Vaccum and with the help of (\ref{inff18}) the normalized solution  reads
\begin{equation}\label{inff41}
\xi_k(t)=\sqrt{\frac{\pi}{4(p-1)}}\sqrt{\frac{1}{a_e\omega_e}}\sqrt{t}\,{\cal H}_{\nu}^{(1)}[x(t)].
\end{equation}
Now using in the IR-sector the asymptotic formula ${\cal H}_{\nu}^{(1)}[x(t)]\simeq \frac{i}{\pi}\Gamma(\nu)[x(t)/2]^{-\nu}$  the equation (\ref{inff19}) leads to
\begin{equation}\label{inff42}
\left<\delta\phi^2\right>=\frac{(-1)^{2\nu}}{2^{3-2\nu}\pi^3(p-1)^{1-2\nu}}\frac{\Gamma^2(\nu)}{\omega_ea^{1-2\nu}_e}t_e^{1-2\nu p}t^{1-q+2(p-1)\nu}\left(\frac{a}{a_e}\right)^{-3}\int_{0}^{\epsilon k_H}\frac{dk}{k}\,k^{3-2\nu}.
\end{equation}
Thus, the squared $\phi$-fluctuations have a power-spectrum ${\cal P}_{\phi}(k)$ given by
\begin{equation}\label{inff43}
{\cal P}_{\phi}(k)\sim k^{3-\frac{\sqrt{1-4\beta}}{p-1}},
\end{equation}
with $\beta$ given by (\ref{inff39}). The spectrum (\ref{inff43}) results nearly scale invariant for $\beta\simeq -\frac{35}{4}-9p^2+18p$. The power-spectrum for the fluctuations of the inflaton field corresponds to the spectral index
\begin{equation}\label{inff44}
n_{\phi}=4-\frac{\sqrt{1-4\beta}}{p-1}.
\end{equation}
It follows from observational data that $n_{\phi}=0.97\pm 0.03$ \cite{Obsd}. It corresponds to $\beta=-0.875-9p^2+18p$ and $\beta=-9.1136-9.3636p^2+18.7272p$, respectively. The first case is achieved when the algebraic equation
\begin{equation}\label{inff45}
\frac{0.00694\omega_e^2\pi(1+\gamma)q^4}{p^3M_p^2t_e^{2q}(q-4)}(594p+27p^2-18q+9qp^2+6q^2p+q^3-108pq-315-324p^2)=\frac{3M_p^2}{4\pi}-\frac{M_p^4t_e^{2q}p}{\pi^3(1+\gamma)^3\omega_e^2q^4},
\end{equation}
is satisfied. It is a 4th degree equation that allow us to fix  $\omega_e$. In the second case we obtain that
\begin{eqnarray}\label{inff46}
&&\frac{0.000011\pi(1+\gamma)\omega_e^2q^4}{p^3M_p^2t_e^{2q}(4-q)}(-2.05\cdot10^5-2.1068\cdot 10^5p^2+3.87\cdot10^5p+16875q^2-11250q+5625qp^2+3750q^2p+625q^3-67500pq)\nonumber\\
&& =\frac{3M_p^2}{4\pi}-\frac{M_p^4t_e^{2q}p}{\pi^3(1+\gamma^3)\omega_e^2q^4},
\end{eqnarray}
must hold for $\omega{e}$. For the electromagnetic modes it follows from the equation (\ref{inff30}) that
\begin{equation}\label{inff47}
\ddot{\zeta}_{\kappa}+\left[\frac{\kappa^2t_{e}^{2p}}{a_e^2t^{2p}}+\frac{3p\left(1-\frac{3}{4}p\right)}{t^2}+\frac{\gamma^2\omega_e}{b^2t_e^q}-\frac{2\gamma^2\omega_e}{b^2t_{e}^{3q/2}}t^{q/2}+\frac{\omega_e\gamma^2}{b^2t_e^{2q}}t^{q}\right]\zeta_{\kappa}=0.
\end{equation}
As the inflationary period is too short, the previous equation can be approximated during inflation by
\begin{equation}\label{inff48}
\ddot{\zeta}_{\kappa}+\left[\frac{\kappa^2t_{e}^{2p}}{a_e^2t^{2p}}+\frac{3p\left(1-\frac{3}{4}p\right)}{t^2}-\frac{2\gamma^2\omega_e}{b^2t_{e}^{3q/2}}t^{q/2}\right]\zeta_{\kappa}=0.
\end{equation}
In general it is not an easy task to solve this equation analitically, however, when $q=-4$ this equation can be written in a simplier form as
\begin{equation}\label{inff49}
\ddot{\zeta}_{\kappa}+\left[\frac{\kappa^2t_e^{2p}}{a_e^2t^{2p}}+\frac{3p\left(1-\frac{3}{4}p\right)-\frac{\sigma}{p}}{t^2}\right]\zeta_{\kappa}=0,
\end{equation}
where
\begin{equation}\label{inff50}
\sigma=\frac{8\pi\gamma^2(1+\gamma)\omega_e^2t_e^{10}}{M_p^2}.
\end{equation}
Hence, according to (\ref{inff32}) the normalized solution of (\ref{inff50}) is
\begin{equation}\label{inff51}
\zeta_{\kappa}(t)=\frac{i}{2}\sqrt{\frac{\pi}{a_e^3(p-1)}}\,\sqrt{t}\,{\cal H}_{\mu}^{(1)}[z(t)],
\end{equation}
where $z(t)=\frac{\kappa t_e^p}{a_e}\frac{t^{1-p}}{p-1}$,  ${\cal H}_{\mu}^{(1)}[z(t)]$ denotes the first kind Hankel function and $\mu=\sqrt{1-12p+9p^2+4\sigma p^{-1}}\,/[2(p-1)]$. \\

Now, employing the formula ${\cal H}_{\mu}^{(1)}\simeq\frac{i}{\pi}\Gamma(\mu)[z(t)/2]^{-\mu}$ the large scale squared fluctuations for the electromagnetic field (\ref{inff33}) are given by
\begin{equation}\label{inff52}
\left<\delta A^2\right>_{IR}=\frac{2^{2\mu-1}}{\pi^3a_e^{3-2\mu}}\frac{\Gamma^2(\mu)}{(p-1)^{1-2\mu}}t_{e}^{(3-2\mu)p}t^{1-3p-2\mu(1-p)}\int_{0}^{\vartheta\kappa_{H}}\frac{d\kappa}{\kappa}\kappa^{3-2\mu}.
\end{equation}
Thus, the squared $A$-fluctutations have a power spectrum ${\cal P}_{A}(\kappa)$ given by
\begin{equation}\label{inff53}
{\cal P}_{A}(\kappa)\sim \kappa^{3-\frac{\sqrt{1-12p+9p^2+4\sigma p^{-1}}}{p-1}}.
\end{equation}
Clearly, the spectrum (\ref{inff53}) is nearly scale invariant for $\sigma =2p-(3/2)p^2$. The spectral index $n_A$ associated for the $A$-fluctuations (\ref{inff52}) is $n_A=4-\frac{\sqrt{1-12p+9p^2+4\sigma p^{-1}}}{p-1}$. From the observational the spectral index takes the values $n=0.97\pm 0.03$. Thus, $n_A$ fits the observations for the spectral index when  $\omega_e^2t_e^{10}=(M_p^2/8\pi\gamma^2(1+\gamma))[(15/4-2n+n^2/4)p-(5-4n+n^2/2)p^2+(7/4-2n+n^2/4)p^3]$ is valid. \\

On the other hand, it is not difficult to verify that the modes (\ref{inff51}) become real on the IR-sector. This happens when
\begin{equation}\label{classic1}
\frac{1}{{\cal N}(t)}\sum_{\kappa=0}^{\kappa\simeq\vartheta \kappa_H}\left|\frac{Im(\zeta_{\kappa})}{Re(\zeta_{\kappa})}\right|\ll 1,
\end{equation}
where ${\cal N}(t)$ is the time dependent number of degrees of freedom during inflation \cite{In7}. Thus for modes inside the interval $10^3\ll \vartheta<\kappa_H/\kappa$ the relation $[\delta A_{\alpha}|_{IR},\dot{\delta A}_{\alpha}|_{IR}]\simeq 0$ holds. Therefore, $\delta A_{\mu}|_{IR}$ can be considered as a classical emergent electromagnetic field and the equation $\bar{B}=\nabla\times\delta \bar{A}|_{IR}$ has sense.

\section{Seeds of Cosmic Magnetic Fields}

\begin{figure}[h]
	\begin{center}
			\centering
		\includegraphics[scale=0.65]{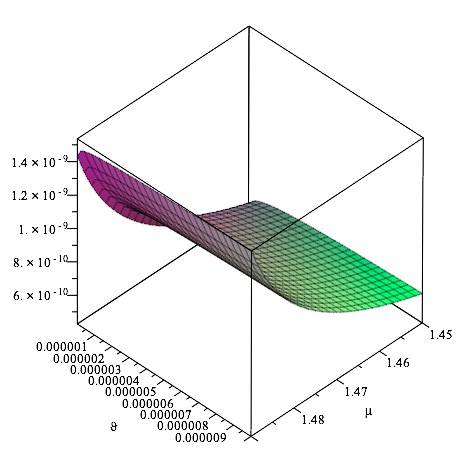}
			\captionsetup{justification=centering, margin=2.4cm}
		\caption{This plot shows $\left<B_{phys}^{(N)}\,^2\right>^{1/2}$, given in Gauss units, versus $\mu_B$ and $\vartheta$. We have used $1.45\leq \mu_B\leq 1.49$ which  correspond to spectral index $1.1\leq n_B\leq 1.02$. The parameter $\vartheta$ varies from $10^{-8}$ to $10^{-5}$ and these values correspond in the present time to the length scales: $10^3-10^{6}$ Mpc. In this plot we have employed  $p=1.000124$.}\label{fig1}
	\end{center}
\end{figure}

Once we have calculated the spectrum for electromagnetic fields during inflation we are in position to obtain the corresponding spectrum for the seeds of cosmic magnetic fields. With this idea in mind it is not difficult to see that according to (\ref{inff22}) the spatial components of the electromagnetic potential obey
\begin{equation}\label{cmf1}
\ddot{\delta A}^{i}+3H\dot{\delta A}^{i}-\frac{1}{a^2}\nabla^2\delta A^{i}+\gamma^2\omega(\phi_c)\phi_c^2\delta A^{i}=0.
\end{equation}
Employing the comoving base $\lbrace E_{0}=e_0,E_{i}=\frac{a_e}{a}e_i\rbrace$ with $\lbrace e_t=\frac{\partial}{\partial t},e_r=\frac{\partial}{\partial r},e_{\theta}=\frac{1}{r}\frac{\partial}{\partial\theta},e_{\phi}=\frac{1}{r sin\theta}\frac{\partial}{\partial\phi}\rbrace$, it follows from (\ref{inff21}) that the comoving magnetic field $\bar{B}_{com}$ satisfy $\nabla\cdot\bar{B}_{com}=0$ \cite{Jackson}. Thus in terms of $\bar{B}_{com}=\nabla\times\delta \bar{A}_{com}$ for a power-law inflation the equation (\ref{cmf1}) reads 
\begin{equation}\label{cmf2}
\ddot{B}^{i}_{com}+H\dot{B}^{i}_{com}-\frac{1}{a^2}\nabla^2B^{i}_{com}+(\gamma^2\omega(\phi_c)\phi_c^2-2H^2-\dot{H})B^{i}_{com}=0.
\end{equation}
Now, analogously to the previos analysis, we can use the Fourier expansion
\begin{equation}\label{cmf3}
B^{i}_{com}(t,\bar{r})=\frac{e^{-\frac{1}{2}\int Hdt}}{(2\pi)^{3/2}}\int d^3\kappa_r \sum_{l=1}^{3}\in^{i}_{(l)}(\kappa_r)\left[d_{\kappa_r}^{(l)}e^{i\bar{\kappa}_r\cdot\bar{r}}\Theta_{\kappa_r}(t)+d_{\kappa_r}^{(l)} \,^{\dagger}e^{-i\bar{\kappa}_r\cdot\bar{r}}\Theta_{\kappa_r}^{*}(t)\right],
\end{equation}
where $d_{\kappa_r}^{(l)} \,^{\dagger}$ and $d_{\kappa_r}^{(l)}$ are the creation and annihilation operators and $\in^{i}_{(l)}(\kappa_r)$ are the $3-$polarisation vectors that obey $\in_{(i)}\cdot\in_{(j)}=h_{ij}$. Thus, the modes $\Theta_{\kappa_r}(t)$ are determined by the equation
\begin{equation}\label{cmf4}
\ddot{\Theta}_{\kappa_r}+\left[\frac{\kappa_r^2t_e^{2p}}{a_e^2t^{2p}}+\frac{\frac{3}{2}p-\frac{9}{4}p^2-\frac{\sigma}{p}}{t^2}\right]\Theta_{\kappa_r}=0.
\end{equation}
With the help of (\ref{inff32}) the normalized solution of (\ref{cmf4}) is given by
\begin{equation}\label{cmf5}
\Theta_{\kappa_r}(t)=\frac{i}{2}\sqrt{\frac{\pi}{a_e^3(p-1)}}\,\sqrt{t}\,{\cal H}_{\mu_B}^{(1)}[Z_{B}(t)],
\end{equation}
where $\mu_B=\sqrt{1-6p+9p^2+4\sigma p^{-1}}/[2(p-1)]$ and $Z_{B}(t)=\frac{\kappa_r t_e^p}{a_e}\frac{t^{1-p}}{p-1}$.
In this manner, the squared $B_{com}$ fluctuations of the seed magnetic field read
\begin{equation}\label{cmf6}
\left<\delta B_{com}^2\right>_{IR}=\frac{\Gamma^2(\mu_B)}{\pi^3 2^{3-2\mu_B}}\frac{t_e^{(1-2\mu_B)p}}{a_e^{3-2\mu_B}(p-1)^{1-2\mu_B}}t^{(1-2\mu_B)(1-p)}\int_{0}^{\vartheta\kappa_{H}}\frac{d\kappa_r}{\kappa_r}\kappa_r^{3-2\mu_B}.
\end{equation}
Therefore, the power spectrum ${\cal P}_B(\kappa_r)$ on cosmological scales is
\begin{equation}\label{cmf7}
{\cal P}_{B}(\kappa_r)=\frac{\Gamma^2(\mu_B)}{\pi^3 2^{3-2\mu_B}}\frac{t_e^{(1-2\mu_B)p}}{a_e^{3-2\mu_B}(p-1)^{1-2\mu_B}}t^{(1-2\mu_B)(1-p)}\kappa_r^{3-2\mu_B}.
\end{equation}
It is not difficult to verify that the spectrum of $\left<\delta B_{com}^2\right>_{IR}$ is nearly scale invariant when $\sigma=2p-3p^2$. The spectral index $n_B=4-\sqrt{1-6p+9p^2+4\sigma p^{-1}}/(p-1)$ is compatible with observational data $n=0.97\pm 0.03$ when $\omega_e^2=\frac{M_p^2}{8\pi\gamma^2(1+\gamma)t_e^{10}}[(15/4-2n+n^2/4)p-(13/2-4n+n^2/2)p^2+(7/4-2n+n^2/4)p^3]$. After integration in (\ref{cmf6}) we obtain for a nearly scale invaraint spectrum on the IR-sector 
\begin{equation}\label{cmf8}
\left<\delta B_{com}^2\right>_{IR}\simeq\frac{\Gamma^2(\mu_B)}{\pi^3 2^{3-2\mu_B}}\frac{t_e^{-2p}\vartheta^{3-2\mu_B}}{(p-1)^{1-2\mu_B}(3-2\mu_B)}t^{2(p-1)},
\end{equation}
which can be written in terms of the Hubble parameter as
\begin{equation}\label{cmf9}
\left<\delta B_{com}^2\right>_{IR}\simeq\frac{\Gamma^2(\mu_B)}{\pi^3 2^{3-2\mu_B}}\frac{\vartheta^{3-2\mu_B}}{p^2(p-1)^{1-2\mu_B}(3-2\mu_B)}\left(\frac{a_e}{a}\right)^{-2p}H^2.
\end{equation}
Now, as it is well-known the physical magnetic field and the comoving one are related via the formula $B_{phys}\sim a^{-2}B_{com}$. Moreover, as after inflation $B_{phys}$ decreases as $a^{-2}$, an estimation of $B_{phys}$ in the present time, here denoted by $B_{phys}^{(N)}$, is given by \cite{In7}
\begin{equation}\label{cmf10}
\left.\left<B_{phys}^{(N)}\,^{2}\right>^{1/2}\right|_{IR}\simeq \left(\frac{a_0}{a_e}\right)^4\left.\left<B_{com}^{2}\right>^{1/2}\right|_{IR}\simeq 10^{-136}\left.\left<B_{com}^{2}\right>^{1/2}\right|_{IR},
\end{equation}
being $a_0=a(t_{now})$ and where we have taken into account that the size of the horizon at the end of inflation is given approximately by $\sim 3.6\cdot 10^{-6}\,cm$ and that the size of the radius of the present observable universe is around $\sim 10^{28} cm$. Finally, in the figure (\ref{fig1}) we show a plot of $<B_{phys}^{(N)}\,^{2}>^{1/2}|_{IR}$ versus $\mu_B$ and $\vartheta$. It can be easily seen from this plot that $<B_{phys}^{(N)}\,^{2}>^{1/2}|_{IR}$ reach values of the order to $\lesssim 10^{-9}\, Gauss$ in general when $p\to 1$. In particular we have taken for this plot $p=1.000124$.

\section{Final Remarks}

In this letter we have derived inflationary cosmological settings from a gravitoelectromagnetic theory. The gravitoelectromagnetic theory has been formulated in the framework of a recent  Weyl-invariant geometrical scalar-tensor theory of gravity in which  the electromagnetic potential and the inflaton scalar field have both geometrical origin, in the sense that they appear in the Weyl-invariant action (\ref{f14})  as part of the affine structure of the space-time. Due to the cosmological principle we start with fluctuations of the electromagnetic potential $\delta A_{\mu}$ only at quantum scales. In this formalism we found that the electromagnetic fields at cosmological scales are generated by means of a classicalization mechanism of $\delta A_{\mu}$ at the end of inflation. The quantum fluctuations for the inflaton field have been studied in the semiclassical approximation. We found that in the decoupling gauge election (\ref{inff20}) the fluctuations of the inflaton and the electromagnetic fields are related. \\

As an application of the formalism we study the case of a power-law inflationary model ($H=p/t,\,\, p>1$), in which we have also assumed a power-law for the $\omega(t)$ function ($\omega(t)\sim t^{q}$). For the quantum fluctuations of the inflaton field we obtain a nearly scale invariant power spectrum for $\beta\simeq -\frac{35}{4}-9p^2+18p$ as is evident in (\ref{inff43}). We have calculated the power spectrum of seed magnetic fields at the end of inflation (\ref{cmf7}), achieving the nearly scale invariance for $\sigma=2p-3p^2$. We obtain that in the present epoch the strength of seed magnetic fields $<B_{phys}^{(N)}\,^{2}>^{1/2}|_{IR}$ reach values of the order $\lesssim 10^{-9}\, Gauss$ when $p\to 1$, in good agreement with CMB observations \cite{CMFOB}. In particular for $p=1.000124$ we show a plot exhibing this behavior. It is important to note that in our formalism the inflaton field can be associated with the geometrical Weyl scalar field.

\section*{Acknowledgements}

\noindent M. Montes acknowledges Centro Universitrio de los Valles of Universidad de Guadalajara for financial support.   J.E.Madriz-Aguilar  acknowledges CONACYT
M\'exico, Centro Universitario de Ciencias Exactas e Ingenierias and Centro Universitario de los Valles of Universidad de Guadalajara for financial support.  V. Granados acknowledges Universidad de Guadalajara for financial support.

\end{document}